\documentclass[conference]{IEEEtran}
\IEEEoverridecommandlockouts
\usepackage{cite}
\usepackage{amsmath,amssymb,amsfonts}
\usepackage{algorithmic}
\usepackage{graphicx}
\usepackage{textcomp}
\usepackage{xcolor}
\def\BibTeX{{\rm B\kern-.05em{\sc i\kern-.025em b}\kern-.08em
    T\kern-.1667em\lower.7ex\hbox{E}\kern-.125emX}}
\begin{document}

\title{Wider: Scale Out Blockchain With Sharding by Account\\
}

\author{\IEEEauthorblockN{1\textsuperscript{st} Jia Kan}
\IEEEauthorblockA{
\textit{Xi'an Jiaotong–Liverpool University}\\
Suzhou, China \\
}
\and
\IEEEauthorblockN{2\textsuperscript{nd} Jie Zhang}
\IEEEauthorblockA{
\textit{Xi'an Jiaotong–Liverpool University}\\
Suzhou, China \\
}
\and
\IEEEauthorblockN{3\textsuperscript{rd} Xin Huang}
\IEEEauthorblockA{
\textit{Taiyuan University of Technology}\\
Taiyuan, China \\
}

}

\maketitle

\begin{abstract}
The development of blockchain applications increased the demand for blockchain performance.
Among the attempts of many blockchain scale-out solutions, sharding can improve performance and reduce the storage requirements of each node.
Sharding enhances the throughput of the entire blockchain.
Most sharding solutions have a fixed number of shards. 
We propose Wider chain with sharding by account.
The number of shards is increased to unlimited.
Meanwhile, Wider combines the idea of rollup and retains the main chain to confirm the status of the subchains.
This design can avoid the Proof of Work computing power being scattered to the subchains and ensure the security of the blockchain.
\end{abstract}


\section{Introduction}

Blockchain enters the era of application.
Use cases have emerged, such as decentralized financial services (DeFi).
The transaction-per-second (TPS) limitation prevents blockchain from being used on a large scale.
The scarcity of TPS makes the transactions more expensive.
The decentralized network is valuable when more people are connected. 
It is very challenging to improve the performance of the blockchain without raising the threshold of node resources.
Once the TPS was increased, the blockchain would still face the problem of data exploding.
It will be difficult for a single personal computer to act as a full node.
Therefore, the blockchain sharding solution is worth studying.

There are two directions of blockchain sharding: BFT-based and PoW-based.
This work follows the PoW-based works such as Monoxide\cite{Monoxide} and OHIE\cite{OHIE}.
Monoxide requires the miners to split the overall computation capacity on different shards.
To avoid the security loss for the divided computation, Monoxide modified the mining rule for multi-shards.
It allows the miner who finds a block to append one block on all the other shards.
OHIE started mining on all the shards, but only append the new block into one of them, randomly decided by the suffix of the new block hash.
The two solutions mentioned above use fixed shards.
A node can choose to host a shard until the requirement of the shard exceeds the node capacity, e.g., disk space or bandwidth.

We propose Wider chain to shard the blockchain by the accounts.
All the transactions for each account are placed on a subchain.
Any transaction is asynchronous and subchains-crossing.
Compared with Layer 2 solutions such as rollup, all the subchains in our solution are globally verifiable.
The consensus algorithm remains as simple as Bitcoin did.
All the computation for blockchain security protects the main chain.
However, it is possible to replace the consensus algorithm with Proof of Staking (PoS) for energy-saving purposes.

\textbf{Our Goal:} This work intends to develop a sharding blockchain solution that satisfies several features.

\begin{itemize}
    \item The higher TPS meets the requirement of blockchain applications but remains decentralized and secure.
    \item Bring down the hard storage requirement on the sharding nodes. As more nodes join the blockchain, each node contributor can use less storage to serve a sharded node.
    \item The overall blockchain on-chain storage will grow with more sharding nodes.
\end{itemize}


\textbf{Contribution and Approach:} We achieve the goals by proposing the one account per shard blockchain data structure in this work.
Instead of putting all transactions in the main chain, we place the transactions on subchains.
The number of shards is up to unlimited instead of fixed shard zones.
The key theory to remove the PoW from the subchain is that \textbf{only when one account for each shard, the consensus is not required since there is no competition}.

Besides, the solution combined sharding by the account with the concept of the rollup.
All the changes on subchains will be confirmed in the main chain block.
Thus, many transactions on a subchain can consume only a tiny space on the main chain.
It results in a higher TPS.
Rollup moves the computation to the off-chain.
Unlike rollup, in Wider, all the subchains are on-chain data, which is publicly verifiable.
Event atomic cross-chain transactions rely on main chain consensus.
So all the computation ensures the main chain security.

\textbf{Result:} We implement the prototype of Wider, which is available in open-source.
Our experiment shows that the single node TPS ranges from 2000 to 5000 per second by adjusting the number of concurrent transaction accounts.


\section{Background}

Internet applications rely on network effects, which connect as many individuals as possible.
Decentralized applications in the blockchain era would still serve massive users.
However, blockchain is still limited in terms of large-scale use.
The mainstream blockchain Bitcoin\cite{Bitcoin} has only 7 TPS, and Ethereum\cite{Ethereum} still has only 15 TPS.
In addition to the low TPS, a single node can hardly carry the entire blockchain data with the growth of historical chain data.
Therefore, sharding will be the inevitable solution for blockchain scaling out.

Bitcoin and Ethereum have very limited TPS.
People try to solve the problem from several angles\cite{ScalabilityofBlockchain}, such as consensus, rollup, sharding.

The earliest attempts were changing consensus.
Bitcoin-NG\cite{BitcoinNG} modified PoW to allow the small blocks produced after the big block.
A miner who solves the PoW puzzle can lead to building small blocks until another miner produces the next big block.
There is no puzzle for the small blocks, so the big block is like a vote.
However, the miner cannot guarantee to keep online after the big block submission.
Algorand\cite{Algorand} replaced PoW computation with randomness algorithm.
Ouroboros\cite{Ouroboros}\cite{Ouroborospraos} adopted PoS direction.

Rollup is one of the most famous concepts for Layer 2.
It includes Optimistic-Rollup and zk-Rollup.
Since many transactions are processed off-chain, only the result is appended on-chain.
The off-chain data are not available for verification. Meanwhile, the blockchain is not allowed to visit the off-chain data.
The Optimistic-Rollup uses fraud-proof against the invalid transaction.
The zk-Rollup relies on advanced zero-knowledge cryptography.

On sharding, many works have been proposed:
Elastico\cite{Elastico}, OmniLedger\cite{OmniLedger} and RapidChain\cite{RapidChain} are BFT-based.
Monoxide\cite{Monoxide} and OHIE\cite{OHIE} remain to PoW design.
Both Monoxide and OHIE use fixed sharding zones to scale the blockchain and achieve outstanding performance.
We proposed a solution Wider with the unlimited shards combined with the rollup concept.

\section{Design}

The Wider chain consists of the main chain and countless subchains.
The main chain consists of blocks, generating new blocks at regular intervals.
Subchains consist of transactions. 
A user can append the transaction to the subchain with the corresponded private key.
Each subchain is a shard, and the shard keeps all transactions of an account.
The main chain blocks do not contain transactions but are used to confirm the latest states of the subchains.
Our sharding solution works with either PoW or PoS algorithm.
We choose PoW here to avoid the token for the consensus, as PoS requires token staking during the consensus computing.
In our sharding solution, the computing power of PoW is concentrated on the main chain.

A user sends a transaction with the three steps in general: 
1. The sender user appends a transaction on the subchain signed with his private key. 
2. The main chain confirms the subchain in the block. 
3. The receiver claims the receiving fund on another subchain by reference to the main chain block and the sending transaction.

The blockchain is desired to run on millions of standard devices with limited storage (up to several TB hard disks for nowadays) and average computation capacity, such as CPU.
Higher TPS will generate a large amount of block history data, so it is difficult for a single node to have sufficient disk space to save all the data.
The storage space on a full node would be consumed fast.
As a decentralized system, the full node of Bitcoin replicates the same ledger data.
It is feasible to bring down the resource requirements for a single node only when nodes cooperate to save data.
In our design, the main chain no longer contains all the transactions.
Nodes can choose to save the part of transaction history data of some shards according to their storage capacity.
Nodes with limited resources can also participate in ledger data preservation and verification.

\subsection{Multi chain data structure}

\begin{figure*}[htbp]
\centerline{\includegraphics[width=.7\paperwidth]{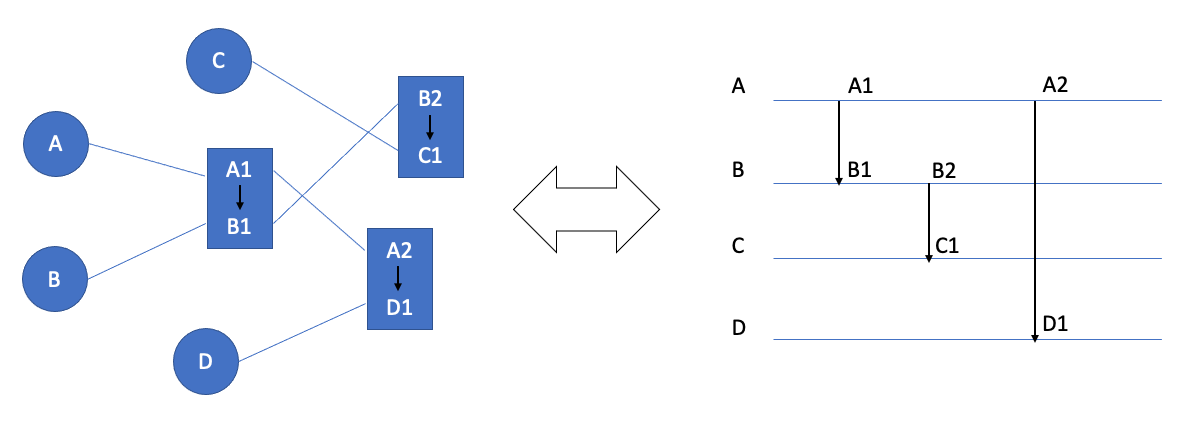}}
\caption{The identical transform from GraphChain data structure into unlimited subchains.}
\label{fig}
\end{figure*}

Monoxide and OHIE are splitting Bitcoin's main chain into fixed shards.
(This would slow down the storage consumption on each node.)
Our solution steps further into the problem, changing from the fixed shards to the unlimited sharding.
The subchains data structure is identically transformed from Kan\cite{Graphchain}, shown in Fig~1.

We model the relationship between no sharding, fixed shards, and unlimited shards.
We define $A$ as the total accounts number of the blockchain, and $S$ as the total number of shards in a blockchain, and $a_{s}$ as the accounts number for the $s^{th}$ shard.
So we have:

$$ A = \sum_{s=1}^{S}{a_s} $$

Alternatively, we can define $\overline{a}$ as the average number of accounts per shard.
So we have $ A = \overline{a} \cdot {S} $.
\begin{itemize}
    \item The Bitcoin and Ethereum are not sharding, which equates to one shard as $S = 1$. Then we have $ A = \overline{a} $.
    \item Blockchain uses fixed number of shards, then we have $S > 1$ as an integer.
    \item Our solution forces $\overline{a} = 1$, so we got $A = S$.
\end{itemize}

The multi-chain data structure is named subchains in our sharding design.
Besides the subchains, the system retains the main chain for all the subchains' security.

\subsection{Mining on main chain}

The blockchain produces a block approximately every ten minutes in the Bitcoin system, and each block contains many transactions.
In Wider, the main chain is equivalent to the blockchain of the Bitcoin system.
Nevertheless, no transactions are placed on the main chain.
The main block confirms subchain status.

Miners received the transaction information from the blockchain broadcast network.
After verification, the transactions were put into the local transaction pool.
The miner runs the PoW algorithm, collects the status of the subchain that needs to be confirmed in the main chain, and then looks for the block hash that satisfies the difficulty.
Our design draws on the concept from rollup: when a user sends many transactions, the transactions on the subchain will be aggregated, and only the final state of the subchain would be marked onto the main chain block.
Therefore, many transactions can be confirmed with less block space in the main chain.

\subsection{Transaction cross subchains}

Each shard in Wider is equal to a subchain.
A subchain represents an account.
All the transactions from an account were put on a subchain.
In our design, we follow the PoW consensus as Bitcoin did.
In addition to the subchains, the main chain is used to confirm the latest status of subchains.

\begin{figure}[htbp]
\centerline{\includegraphics[width=\linewidth]{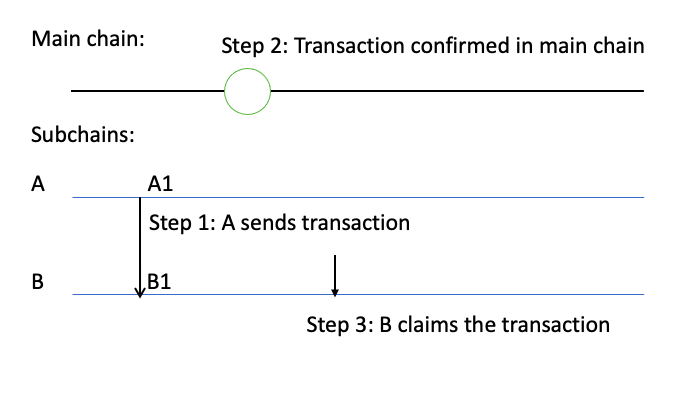}}
\caption{The steps for cross subchains transaction.}
\label{fig}
\end{figure}

There are two situations for the transaction in fixed sharding: intra-shard and cross-shard transfer.
In our solution sharding by account, all the transactions are cross-shard transfers unless a user transfers to himself.
Fig~2 shows the steps for sending transactions and receiving funds.

\begin{table}[htbp]
\caption{Sending transaction data structure}
\begin{center}
\begin{tabular}{|c|c|}
\hline
\textbf{\textit{Data field name}}& \textbf{\textit{Data type}} \\
\hline
Transaction hash & bytes \\
\hline
Parent hash & bytes \\
\hline
Current height & unsigned int \\
\hline
Current address & bytes \\
\hline
Recipient address & bytes \\
\hline
Payment amount & unsigned int \\
\hline
Timestamp & unsigned int \\
\hline
Signature & bytes \\
\hline
\end{tabular}
\label{tab1}
\end{center}
\end{table}

\begin{table}[htbp]
\caption{Receiving transaction data structure}
\begin{center}
\begin{tabular}{|c|c|c|}
\hline
\textbf{\textit{Data field name}}& \textbf{\textit{Data type}} \\
\hline
Transaction hash & bytes \\
\hline
Parent hash & bytes \\
\hline
Current height & unsigned int  \\
\hline
Current address & bytes \\
\hline
Sender address & bytes \\
\hline
Sender transaction hash & bytes \\
\hline
Main chain block hash & bytes \\
\hline
Payment amount to claim & unsigned int \\
\hline
Timestamp & unsigned int \\
\hline
Signature & bytes \\
\hline
\end{tabular}
\label{tab2}
\end{center}
\end{table}

A user sends a transaction with the detail steps:
The account user queries the latest status on the subchain, including available balance and latest transaction height.
The user builds the sending transaction data structure (Table~I), which includes the hash of the parent transaction, the current height, the recipient address, the payment amount, and the timestamp.
The user signs the transaction with his private key.
Then the user calculates the hash value of the above data to get the hash value of the transaction.
Next, the signed transaction will be broadcast on the blockchain network.
The blockchain nodes receive and verify the transaction, then put the transaction into the transaction pool.
The payment amount can not exceed the available balance. 
Otherwise, the transaction will fail during the verification.
Finally, the miners run the PoW algorithm to build a new block on the main chain.
The new block will include the information of updated subchains, which confirms the recent transactions in subchains.

This confirmation of subchains in the main chain is similar to the concept of the rollup.
Compared to rollup, all subchains in Wider are verifiable.
Like Monoxide and OHIE, we categorized Wider as a Layer 1 sharding solution.
All the computation of PoW is to protect the security of main chain.

\subsection{Transaction fund claim}

Once the transaction is sent from an account, the nodes and miners will verify and confirm the transaction on the main chain.
The transaction confirmation has two steps: the main chain conformation and subchain confirmation.
The main chain confirmation ensures the sending action exists and is correct.
Since all the miners are working on the main chain to find a new block, the miners verified the updated subchains.
Mining is the guarantee for transaction security.

Receiving transaction (Table~II) claims the transaction fund.
There is no requirement to perform receiving transactions immediately after main chain conformation.
The subchain may append the receiving transaction before spending the balance.
It is the action to collect the fund.
The receiving transaction would increase the account's balance.
We recommended appending this receiving transaction right before sending the fund next time.

The concurrent number of accounts sending transactions is called width.
A block with a limited size can only confirm a limited number of subchains each time.
When lots of accounts send transactions simultaneously, the users must wait for several blocks to finish all the confirmations.
Thus, the user collecting the received funds and sending new transactions in a batch could avoid unnecessary subchain confirmation.
It would save the main chain block space and increase the TPS.

\subsection{Event atomic}

In Bitcoin, a block contains many transactions.
The transaction is synchronized atomic, which deducts the sender's balance and increases the receiver's balance at the exact moment (assumed no transaction fee). 
In our sharding solution, the transaction is asynchronous atomic.
The sender reduces the balance, and the main chain confirms the transactions on the shard.
The receiver may choose to claim the transaction any time after the main chain is confirmed the subchain.
In Monoxide, such asynchronous atomic of the account balance is defined as event atomic.
We follow Monoxide's definition.

The send actions on different subchains are independent.
The account with enough balance can perform sending action.
In OHIE and Bitcoin, the order of blocks and transactions inside the block is critical.
Swapping the order of blocks or transactions will lead to verification failure, e.g., an account would fail to pay before receiving funds.
Wider does not compare the timestamp of the transactions on the different subchains.
However, within the shard, the transactions are chained with continual height.
When sending funds from one account to another, it checks if the current account has enough balance to payout.
The sending transaction deducts the balance of the account.
The receiving transaction can increase the balance by referencing the main chain block and the send transaction.

\subsection{Security of subchains}

In Wider, there is no consensus algorithm targeting subchains.
The subchain allows the user to fork the chain before the confirmed block is generated.
On the other hand, the main chain is secure enough to freeze the confirmed subchains, as all the computations were used for PoW consensus on the main chain.

Another concern is that a malicious user may try to replace the content of the transaction but remain the transaction hash.
Most of the data fields in the transaction are fixed, such as the parent transaction hash.
The height must be continual.
The probability of finding a hash of different content is negligible, especially when the modification will also lead to transaction signature changes.
The nodes which keep the subchain data will be aware when processing a different transaction with duplicated hash.


\section{Implementation}

The implementation followed by design must consider several aspects.
The nodes should host the shards in redundancy.
Although the number of blockchain nodes is flexible, it is crucial to preserve all shards to avoid blockchain data loss.

Bitcoin and Ethereum only provide full nodes.
Monoxide and OHIE use fixed shards, so the new node can easily choose the shard of fewer nodes to join.
In Wider, there are unlimited shards.
A node will host many shards.
It requires the nodes organized to provide shards redundancy.

The second goal is to speed up the data access among the nodes.
A sharding blockchain splits users' data on different nodes.
The new block generation and verification rely on the data access to the other shards.
A pure Peer to Peer (P2P) network running behind thousands of family firewalls is hard to communicate globally efficiently.

\subsection{Roles}

As we mentioned above, there are two roles in the Wider chain same as Bitcoin: the node and the miner.
The node hosts the blockchain data, and the miner provides blockchain security.
Both the node and miner verify the blockchain transaction data.

Unlike Bitcoin, the node does not keep all the subchains in a sharding blockchain.
Each node keeps the complete main chain and a subset of subchains.
The node is nice to have a public IP address and opened port with enough bandwidth.
Alternatively, at least, it can be accessible from the Internet.
The node is a part of the backbone broadcast network.
When a new block broadcasting arrives on the node, the node checks the subchains confirmed in the block content.
Compared with a full node, the sharding node only needs to verify the incremental subchains kept in the local, which would save much computation in signature and hash verification.

\begin{figure}[htbp]
\centerline{\includegraphics[width=0.8\linewidth]{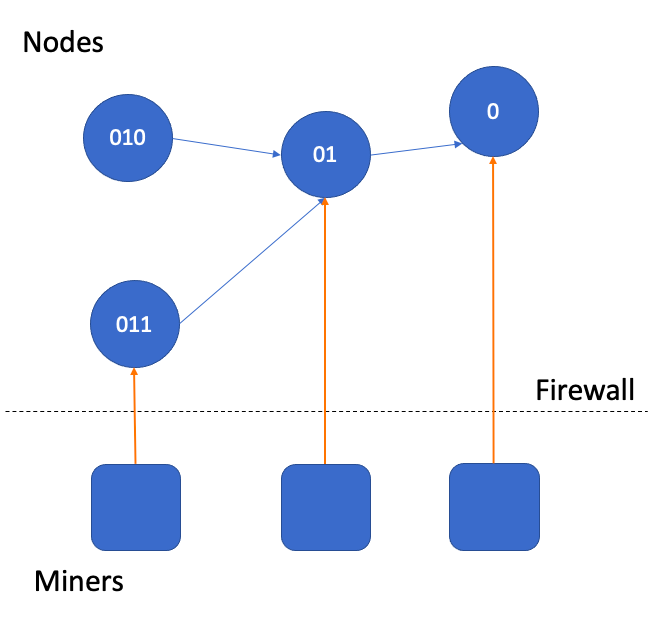}}
\caption{The blockchain roles: nodes and miners.}
\label{fig}
\end{figure}

The miner who works on the PoW algorithm will connect to at least one node, shown in Fig~3.
It listens and waits for the new block or transaction from the broadcast network.
When a new block arrives, the miner marks the updated subchains status.
After subchain status is confirmed, the receiver can claim funds on another subchain.
When a new transaction arrives, the miner connects to the node that hosts the shard of the transaction sender.
The miner downloads the related subchains for verification.
The verified transactions are put in the pool, so the miner builds the next block with the pool transaction by running the PoW algorithm.

\subsection{Network}

\begin{figure*}[htbp]
\centerline{\includegraphics[width=0.7\paperwidth]{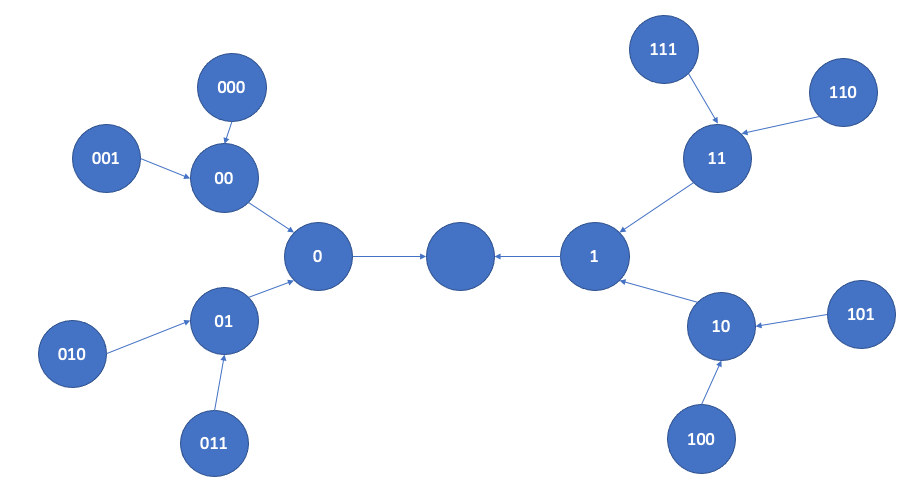}}
\caption{The backbone network for the nodes.}
\label{fig}
\end{figure*}

Either DHT network\cite{Kademlia} or the tree network can be used to build the blockchain broadcast network.
We choose to use the tree network as the backbone network for the sharding nodes.

The DHT network is widely used in P2P transportation.
It is ideal for a decentralized system.
The DHT network node self-assigns the node identity and communicates with the neighborhood.
A node can find others without a central name service.
For the broadcast task, the DHT needs to work with gossip protocol\cite{Gossip}.
A node receives a message and sends it to another contact until all the contacts he knows already hear the message.
In our situation, the shards can be kept by nodes whose identity is close to the shard user address.
However, when the number of nodes is not enough, some shards would get lost.

Fig~4 shows the tree network as the blockchain backbone network.
Compared to the DHT network, the tree network is better organized.
The efficiency structure makes the network better in message broadcast.
The disadvantage of a tree network is the single node failure issue.
This problem can be addressed by adding connections between nearby nodes.

\subsection{Nodes}

\begin{figure}[htbp]
\centerline{\includegraphics[width=0.8\linewidth]{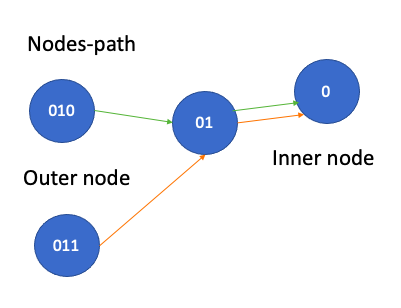}}
\caption{Set of nodes hosting the same shard as the nodes-path.}
\label{fig}
\end{figure}

In the tree network, the connected nodes form a backbone network.
The backbone network is used for broadcasting and placing sharding data.
The node with public access is allowed to join the tree network by appending as a leaf of the tree.

Fig~5 shows the nodes-path for a given account.
The nodes-path is the nodes that host the same shards.
Any account can find the corresponded nodes-path by transforming the account address into binary and truncating the prefix.
When only one node exists, the node is responsible for keeping all the shards.
As more nodes join the tree network, they are assigned to host different shards.
The inner nodes host more shards than the outer in the nodes-path.
A user can find the corresponding nodes-path by his account address.

\subsection{Verification}

The rollup technique is used in Wider.
In theory, the subchain can append arbitrary transactions and wait for confirmation on the following main block.
While the TPS can be easily raised up, the pressure moves on the verification.

Both the node and the miner are required to perform the verification.
The node accepts the new block and checks the subchains' correctness.
The miner verifies the transactions before starting mining the next block.
When there are many transactions to confirm on a subchain, it may cost more computation in verification.

Once the related subchains are fetched to local, more CPU cores or distributed computation can be used for verification.
Parallel computing may reduce the time in processing transactions, which increases TPS.

State transform function (STF) is the concept from the state machine.
The blockchain is modeled as the state machine.
The chain's state transfers to a new state with the given input.
The STF defines the rule to compute the new state from the current state and input.
STF can be used when adding a new transaction or new block.
During the verification, the STF checks the illegal input.

Each subchain will broadcast the latest transaction to get the subchains verified by all the miners and nodes.
The miners received the transactions and put them into the transaction pool.
Before the miner runs a consensus algorithm to find the new block, they must verify the new appending transaction on the subchains.
However, the miner only needs to download a part of the subchain if the subchain has been confirmed.

\section{Experiment}

There are several parameters to measure the Wider blockchain performance:
\begin{itemize}
    \item Transaction width: the number of accounts that participate during the block interval.
    \item Average transactions: the average number of transactions on a subchain.
    \item Block size: the size of the main chain block, which indicates the number of subchains to confirm in a block.
    \item Block interval: the interval of main chain blocks.
\end{itemize}

We implement our sharding blockchain with python 3.8.
The overall performance with python is not as fast as C++ or rust.
Therefore, Wider may achieve better performance once we turn the solution industry-ready.
We run the experiments on Ubuntu 20.04 LTS operation system. 
The test CPU is AMD Ryzen 7 5800 @3.4GHz with 8 Cores.
The blockchain uses the RocksDB database on an SSD hard drive.
For the network, we choose the tree network to achieve better performance in broadcast communication.

\subsection{Blockchain TPS}

We measure the blockchain by adjusting the blockchain width and block size parameters.
As the rollup technique is applied, we know all the subchains updating would be confirmed in the following main chain if the block size is unlimited. 
Our experiment showed Wider can achieve 6000+ TPS by removing the block size limit.
However, there is always a fixed bandwidth in practice so that the block size cannot increase forever.
We also measure different tiers of block size by different block intervals in our experiments.

Followed by Bitcoin and Ethereum, we fix the interval of blocks to 10 minutes and 15 seconds and use only one core of CPU.
So in the first experiment, we set the transaction width from 5,000 to 15,000 accounts for the 10 minutes block interval and 1MB block size and set the transaction width from 100 to 1,000 accounts for the 15 seconds block interval and 40KB block size.

Table~III shows the result:
under the Bitcoin-like setting (10 minutes block interval and 1MB in block size), the sharding blockchain node can achieve 2000-5000 TPS.
Under the Ethereum-like setting (15 seconds block interval and 40KB in block size), the sharding blockchain node can achieve 1000-3000 TPS.

\begin{table}[htbp]
\caption{Receiving transaction data structure}
\begin{center}
\begin{tabular}{|c|c|c|c|}
\hline
\textbf{\textit{TX Width}}& \textbf{\textit{Block size}}& \textbf{\textit{Block Interval}}& \textbf{\textit{TPS}} \\
\hline
100 & 40KB & 15s & 2362 \\
\hline
200 & 40KB & 15s & 2409 \\
\hline
300 & 40KB & 15s & 2026 \\
\hline
400 & 40KB & 15s & 2423 \\
\hline
500 & 40KB & 15s & 1903 \\
\hline
600 & 40KB & 15s & 3276 \\
\hline
700 & 40KB & 15s & 2491 \\
\hline
800 & 40KB & 15s & 2282 \\
\hline
900 & 40KB & 15s & 1656 \\
\hline
1000 & 40KB & 15s & 1461 \\
\hline
5000 & 1MB & 600s & 4937 \\
\hline
6000 & 1MB & 600s & 5407 \\
\hline
7000 & 1MB & 600s & 2293 \\
\hline
8000 & 1MB & 600s & 3692 \\
\hline
9000 & 1MB & 600s & 2906 \\
\hline
10000 & 1MB & 600s & 2596 \\
\hline
11000 & 1MB & 600s & 2502 \\
\hline
12000 & 1MB & 600s & 2497 \\
\hline
13000 & 1MB & 600s & 2313 \\
\hline
14000 & 1MB & 600s & 2233 \\
\hline
15000 & 1MB & 600s & 2159 \\
\hline
\end{tabular}
\label{tab2}
\end{center}
\end{table}

\subsection{Transactions verification}

\begin{figure}[htbp]
\centerline{\includegraphics[width=0.8\linewidth]{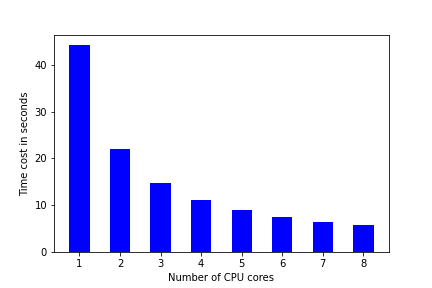}}
\caption{The time cost for transaction verification with more CPU cores.}
\label{fig}
\end{figure}

As the TPS increased, the system bottleneck is now in verification, which requires much more computation.
A single CPU core can perform limited times of the signature (ECDSA) verification operations per second. 

Both the node and the miner are responsible for verifying data after receiving the block or the transactions.
The miner needs to verify all the broadcast network transactions to maximize his incoming.
The node would verify the corresponded subchains after the main chain block confirms.
Multicores CPU can effectually reduce the time for transaction verification.

We experiment on the 10,000 transactions verification with the different number of cores.
Fig~6 shows the time cost in verification.
Thus, the transaction processing speed can rise with the help of other cores.
The computation can even be out-sourcing to a local cluster.
\subsection{Sharding and full node}

\begin{figure}[htbp]
\centerline{\includegraphics[width=0.8\linewidth]{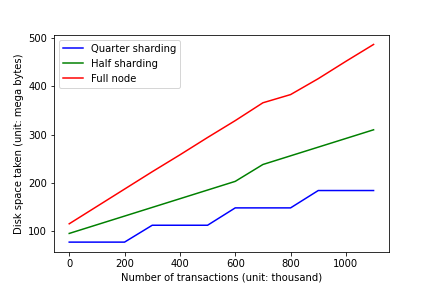}}
\caption{The storage space cost for the full node and the sharding nodes.}
\label{fig}
\end{figure}

The sharding node shall suffer less pressure than a full node in Wider, including CPU and disk usage. 
With more and more nodes joining the blockchain backbone network, the new node is appended to the tree network as the outer leaf.
As there are two leaves of a parent, the leaf node only keeps half of the parent shards and the main chain.

We measure the storage space taken by the RocksDB database for full node, half sharding, and quarter sharding.
The result is shown in Fig~7. 
The sharding node uses less storage space than the full node.

Unlike the Layer 2 or off-chain solutions, all subchain transactions are publicly verifiable.

\section{Discussion}

\subsection{Network latency}

The BFT solution requires the sufficiency bandwidth and the minimized latency for the consensus.
Wider requires the bandwidth for the transactions messages spreading.
The delay in the message would not affect the TPS or performance.
On the other hand, a significant network latency would delay the verification.
As the sharding node does not contain the complete information for verification, it fetches other subchains fragments on demand.
Overall, the latency has less impact on a PoW-based sharding solution than the BFT-based.

\subsection{Incremental data fetch}

Wider remains the main chain on all the nodes.
Every node must keep the main chain synced.
The main chain contains the updated information of the subchains' status.
A subchain updates its status A and B in the main chain at block heights 3 and 5.
When a node receives block 3, it downloads the subchain transactions before A point and gets the subchain verified.
When a node receives block 5, the node verifies the transactions between points A and B.
Thus, a node does not download the full subchain but just a fragment.
The main chain reduces the communication overhead.

\subsection{Public access of node}

The blockchain was famous as the typical P2P application. 
However, the node running behind a firewall connected to a P2P network can hardly achieve high performance.
The Network Address Translation (NAT) protocol on the home router devices even banned the P2P incoming requests.
Towards the blockchain applications, the performance and scale-out should be considered before censorship-resistant with P2P networks.
Indeed, censorship-resistant should be handled with cryptography.

In Wider, we allow the resource contributor to run behind the firewall.
However, the nodes must be publicly accessible from the Internet.
The node is usually located in the data center with good bandwidth.
If the miners want to exchange data, the nodes are the backbone network for message propagation, nodes-path, and proxy.

\subsection{Node failure for broadcast}

The main difference between the DHT network and tree network\cite{Boost} is the efficiency of message broadcasting.
DHT network\cite{Kademlia} uses gossip protocol\cite{Gossip}, which is robust against node failures.
For the performance and organized sharding structure, we chose the tree network.
However, it is vital to catch the message propagation issue caused by node failure.

There is a simple solution to resilience the node failure.
Each node joins the network by connecting to a parent node.
The node should immediately memorize the nearby n-th neighbors' address.
When the node receives a message, it passes the message to the parent and the children first through. Then it also passes the message to all the nearby nodes.
Thus, even if a single node fails, the message will still reach the whole network through the redundancy paths.

\subsection{No consensus on subchain}

Wider remains the main chain.
There is no PoW consensus algorithm targeting for the subchains.
It is essential to understand that a subchain cannot remove the consensus when the number of accounts on a shard is more than one.
As more accounts are placed on the same shard, each subchain must handle the order of transactions.
Since the consensus algorithm is used to get multi parties to agree on the same, the transaction order must be decided with consensus.

In Wider, we make one account one shard.
Thus, the subchains drop the consensus protection.
The security computation focuses on the main chain.
However, the cryptographic hash still keeps the subchain unchangeable.
On the other hand, by removing the consensus on subchains, mechanisms like Chu-ko-nu mining from Monoxide\cite{Monoxide} can be avoided.

\section{Related Work}

There are lots of research works on blockchain performance or the scale-out.
Sharding is one of the promising directions, as it solves the data exploding after performance raise.
However, other attempts have been researched from different angles.
We can roughly categorize the solutions into Layer 1 and Layer 2\cite{ScalabilityofBlockchain}.

The Layer 1 solutions are on-chain.
It focuses on improving the consensus\cite{ConsensusAlgorithms}, data structure, and the network.
SegWit is an early proposal from the BIP (Bitcoin Improvement Proposals), which slightly increased the Bitcoin TPS.
Bitcoin Cash\cite{BitcoinCash} increased the block size up to 8MB.

Bitcoin-NG\cite{BitcoinNG} tries to modify the consensus and incentive rule to achieve higher TPS.
Except for PoW, Proof of Stake (PoS) is well studied.
Ouroboros\cite{Ouroboros}\cite{Ouroborospraos} elects leader in each epoch.
Algorand\cite{Algorand} uses randomness selecting miner.
Besides the cryptography, PoS consensus relies on the token staking to resilience Sybil-Attack.

Changing the data structure is another interesting approach.
Since most blockchain systems are hash chain-based, Directed Acyclic Graph (DAG) allows transactions processed in parallel.
DAGCoin\cite{Dagcoin}, Byteball\cite{Byteball}, Nano\cite{Nano} and IoTA\cite{Iota} are the early DAG adopters.
Phantom\cite{Phantom} applies blockDAG to achieve higher throughput.
Conflux\cite{Conflux} improves DAG with different weights on the edges.
Avalanche\cite{Avalanche} uses DAG but improves the consensus as well.

Sharding is a viable approach to increase performance while reducing the node storage requirement.
Elastico\cite{Elastico}, OmniLedger\cite{OmniLedger} and RapidChain\cite{RapidChain} are BFT-based.
Monoxide\cite{Monoxide} and OHIE\cite{OHIE} remain to PoW consensus.

The Layer 2 solutions are off-chain.
The approaches include payment channel, sidechain, cross-chain, and rollup.

Lightning\cite{Lightning} is the payment channel for Bitcoin, and  Raiden\cite{Raiden} is similar to Lightning but for Ethereum.
Raiden supports ERC20 tokens.

The sidechain is the concept that a parent chain transfers the asset to the child chain and continues to perform operations.
Plasma\cite{Plasma} bridges the asset from an Ethereum smart contract to a sidechain.
Users can withdraw the asset on the main chain at any time.

Cross-chain uses a relay technique connecting many blockchains.
The blockchains are interoperability.
Polkadot\cite{Polkadot} and Cosmos\cite{Cosmos} show the application blockchains work together to achieve overall high performance.

Rollup is famous for Optimistic Rollup and zero-knowledge Rollup.
Arbitrum\cite{Arbitrum} is the Optimistic Rollup speeding up Ethereum.
Meanwhile, the zk-Rollup\cite{zkevm} solution is still under construction.
Besides, Truebit\cite{Truebit} offers the off-chain computation through verifiable computation.

\section{Conclusion}

We present Wider, a sharding blockchain solution with a novel one-shard per account concept.
The new subchains data structure combined with the rollup achieved high performance and reduced the node storage requirement.
The main chain block was used to confirm subchains instead of transactions.
The PoW protects the main chain as secure as Bitcoin.

\section{Future work}

The data structure and solution proposed in this paper are suitable for high-performance blockchain transactions.
In order to run smart contracts in such blockchain data structures, we need to continue research and come up with corresponding theories.
To this end, we are building an EVM-compatible virtual machine and trying to modify it so that the virtual machine can run safely in the multi-chains environment.





\end{document}